\documentclass[3p,11pt]{amsart}
\usepackage{etex}
\usepackage[margin=1.25in]{geometry}
\usepackage[latin1]{inputenc}
\usepackage{amsmath}
\usepackage{stackengine}
\usepackage{a msthm}
\usepackage[alphabetic]{amsrefs}
\usepackage{amsfonts}
\usepackage[all,cmtip]{xy}
\xyoption{rotate}
\usepackage{amsfonts}
\usepackage{amssymb}
\usepackage{amscd}
\usepackage{bbm}
\usepackage{pictexwd,dcpic}
\usepackage[T1]{fontenc}
\usepackage{lmodern}
\usepackage{setspace}
\usepackage{enumerate,multicol}
\usepackage[svgnames]{xcolor}
\usepackage[colorlinks=true, citecolor=Dark Green, linkcolor=Maroon, urlcolor=Purple]{hyperref}
\usepackage{tikz}

\def\0{\mathop{\mathbf{0}_{}}\nolimits}
\def\1{\mathop{\mathbf{1}_{}}\nolimits}

\def\FPR{\text{FPR}}

\def\RR{\text{RR}}

\def\eqgap{.2ex}
\def\overgap{.4ex}
\def\inferrulerule{.2pt}

\newlength\rulelength
\newlength\toplength
\newlength\bottomlength

\newcommand\myinferrule[2]{%
  \stackMath%
  \setlength\bottomlength{\widthof{$#1$}}%
  \setlength\toplength{\widthof{$#2$}}%
  \ifdim\toplength>\bottomlength%
    \setlength\rulelength{\the\toplength}%
  \else%
    \setlength\rulelength{\the\bottomlength}%
  \fi%
  \mathrel{%
    \stackunder[\overgap]{%
      \stackon[\overgap]{%
        \stackanchor[\eqgap]%
          {\rule{\the\rulelength}{\inferrulerule}}%
        {\rule{\the\rulelength}{\inferrulerule}}%
      }{#2}%
    }{#1}%
  }%
}

\theoremstyle{definition}

\title[Reproducibility crisis]{Why ``Redefining Statistical Significance'' Will Not Improve Reproducibility and Could Make the Replication Crisis Worse}

\author[Harry Crane]{Harry Crane}
\address{Department
of Statistics \& Biostatistics, 
 Rutgers University, 
110 Frelinghuysen Road, 
Piscataway, NJ 08854, USA\\
http://www.harrycrane.com\\
Copyright 2017 Harry Crane}
\email{hcrane@stat.rutgers.edu}

\begin{document}
\maketitle
\begin{abstract}
A recent proposal to ``redefine statistical significance'' (Benjamin, et al.\  {\em Nature Human Behaviour}, 2017) claims that false positive rates ``would immediately improve'' by factors greater than two and replication rates would double simply by changing the conventional cutoff for `statistical significance' from $P<0.05$ to $P<0.005$.
I analyze the veracity of these claims, focusing especially on how Benjamin, et al neglect the effects of P-hacking in assessing the impact of their proposal.
My analysis shows that once P-hacking is accounted for the perceived benefits of the lower threshold all but disappear, prompting two main conclusions:
\begin{itemize}
	\item[(i)] The claimed improvements to false positive rate and replication rate in Benjamin, et al (2017) are exaggerated and misleading. 
	\item[(ii)] There are plausible scenarios under which the lower cutoff will make the replication crisis worse. 
\end{itemize}
\end{abstract}

\section{Introduction}\label{section:introduction}

The proposal to ``redefine statistical significance'' \cite{RSS} (henceforth abbreviated as the `RSS proposal' or simply `RSS') is intended to counteract the so-called `reproducibility crisis', i.e., the disproportionate fraction of statistically significant scientific findings that cannot be replicated by subsequent studies.
In championing their proposal, the signatories of RSS claim, ``This simple step would immediately improve the reproducibility of scientific research in many fields.''
The authors go on to assert that ``false positive rates would typically fall by factors greater than two'' and suggest that replication rates would roughly double under their proposal.
 
 In ignoring the effects of P-hacking on false positive rate and replication rate, RSS misses the whole point of the reproducibility crisis.
  By appealing to the same formal technique and empirical evidence \cite{PsychRep} used to support the RSS proposal, I will unmask major conceptual and technical flaws in the RSS argument.\footnote{The analysis in Section \ref{section:effects} is based on the replication study in \cite{PsychRep}, which was conducted on a sample of 97 results in psychology.  I do not analyze the results in \cite{Camerer2016}, which performs a similar study for findings in experimental economics but for a much smaller sample of 18 studies.} 
The analysis presented here is not a counterproposal to RSS, but rather a refutation which is intended to elucidate the proposal's flaws and therefore neutralize the potential damage which would result from its implementation.\footnote{The analysis below focuses on the potential impact of the proposal concerning statistical `significance'.  The RSS proposal also includes a recommendation to regard findings with $0.005<P<0.05$ as `suggestive'.  The motivation for this suggestion, its justification, and perceived impact are not clearly articulated in \cite{RSS}, and thus will not be discussed further here.}
Our analysis may be seen as complementary to, but should not be read in any way as an endorsement of, the critiques and alternative proposals by other authors \cite{Amrhein2017, Lakens2017,McShane2017,Trafimow2017}.
I discuss this last point further in Section \ref{section:concluding}.

\subsection*{Brief summary}

P-hacking and the reproducibility crisis: like smoking and lung cancer, one cannot be discussed without the other \cite{GelmanLokin2014,Head2015,ASA}.\footnote{For simplicity, I use the term `P-hacking' to refer to any unsound statistical practice used in justifying a scientific finding by a significant P-value, including but not limited to ``cherry-picking, P-hacking, hunting for significance, selective reporting, multiple testing and other biasing selection effects'' \cite{Mayo2017}.  
Most salient for our purpose here is that P-values obtained by P-hacking do not warrant the same interpretation as the standard theory assumes.
To be clear, I do not use the term `P-hacking' in a pejorative sense.  For the purposes it is employed here, P-hacking need not be intentional or done with malicious intent.  What is important, however, is that P-hacking cannot be ignored when studying the effect of the RSS and other proposals on reproducibility.
}  Yet the RSS analysis does just that.  Because they do not intend their proposal to combat P-hacking directly, the advocates of RSS seem to think that they can set it aside.  
 But in ignoring the effects of P-hacking, RSS arrive at overly optimistic projections and tout misleading conclusions about the ``immediate'' benefits of their proposal.
With P-hacking accounted for, we arrive at much more realistic, and sobering, conclusions about the potential impact of redefining statistical significance.

  \begin{figure}[t]
\begin{center}
\scalebox{.5}{\includegraphics{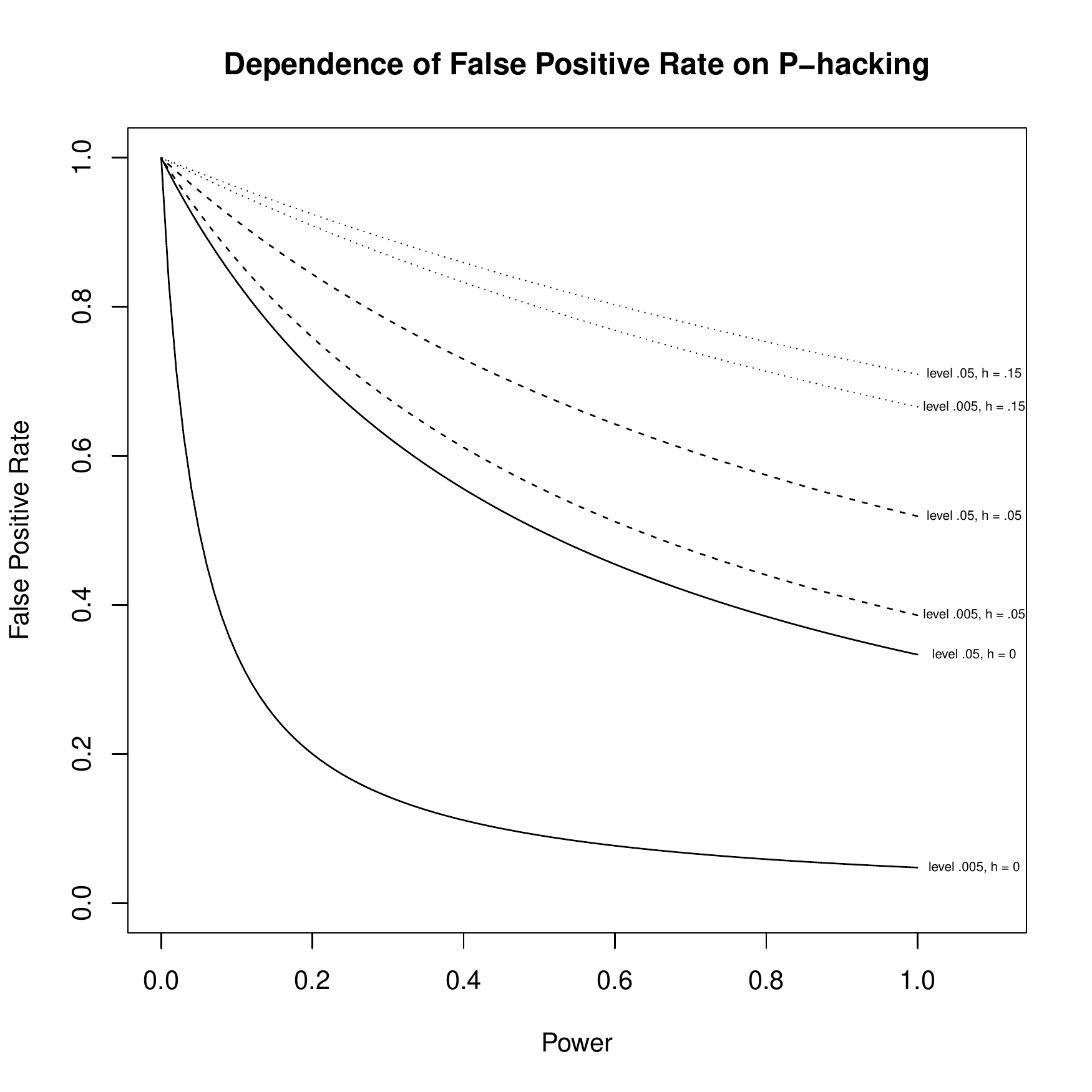}}
\end{center}
\caption{False positive rate \eqref{eq:FPR} for different significance levels ($\alpha=0.05, 0.005$) and hacking rates ($h=0,0.05,0.15$).  Solid lines correspond to false positive rate without P-hacking; dashed lines to FPR with $h=0.05$ (i.e., 5\% of all P-values are hacked); and dotted lines to FPR with $h=0.15$.}
\label{fig:P-hacking}
\end{figure}

To foreshadow the analysis presented in Section \ref{section:effects}, Figure \ref{fig:P-hacking} plots the false positive rate (FPR) versus statistical power for different combinations of significance level and P-hacking rate (i.e., the proportion of all P-values obtained by P-hacking).  The solid lines, which correspond to the traditional false positive rate in the absence of P-hacking (see Equation \eqref{eq:FPR-nohack} below), are also shown as part of \cite[Figure 2]{RSS}.
These lines suggest a substantial improvement to FPR under the reduced significance level: for statistical power of 0.80, false positive rate is projected to decrease from 0.38 to 0.06.
  But if 15\% of all P-values are hacked,\footnote{The values of $h=0.05$ and $h=0.15$ in Figure \ref{fig:P-hacking} are chosen as the extreme estimates obtained in Section \ref{section:effects}, which based on the replication study in \cite{PsychRep} suggests that between 5\% and 15\% of all P-values are hacked.} then the false positive rate would decrease from 0.75 to 0.71, just a 5\% improvement, as a result of the lower cutoff.  And even at the low end of our estimate for the prevalence of P-hacking (i.e., $h=0.05$), the false positive rate would only improve from 0.57 to 0.44.   Regardless of how much the false positive rate improves in relative terms, the end result (44\%-71\% false positives) is hardly worth celebrating: the false positive rate is bound to remain much higher than the suggested 0.06 level presented in \cite{RSS}.  In direct conflict to the main argument in \cite{RSS}, Figure \ref{fig:P-hacking} illustrates in stark terms the deception lurking in the RSS argument and makes clear that decreasing the significance level will hardly make a dent in the reproducibility crisis.
 
  The forthcoming analysis develops the framework from which Figure \ref{fig:P-hacking} is derived.
Though the analysis is set in the context of the RSS proposal, the principles underlying it are relevant beyond the specific numbers (i.e., 0.05 versus 0.005) or methods under consideration (e.g., hypothesis testing, Bayes factors, confidence intervals, etc.).  P-hacking and other forms of statistical misuse and malpractice are endemic to science at all levels.  Its role in the reproducibility crisis cannot be ignored, especially when evaluating proposals such as the one considered here.

\section{Null Hypothesis Significance Testing (NHST)}\label{section:NHST}\label{section:nohack}

Under the Null Hypothesis Significance Testing (NHST) paradigm, a null hypothesis ($H_0$) is tested against an alternative hypothesis ($H_1$) by computing a P-value, defined as the probability (under a true null hypothesis) that a certain test statistic attains a value as or more extreme than what is actually observed.
Small P-values, which correspond to observations that are unlikely to have occurred if the null hypothesis were true, are interpreted as evidence against $H_0$.
In fields adhering to the NHST paradigm, it is conventional to `reject $H_0$' and confer the label of `statistical significance' whenever a P-value falls below a pre-described threshold $0<\alpha<1$.  

\begin{table}[t]
\begin{tabular}{|c||c|c|}
\hline
& $H_0$ true & $H_0$ false\\\hline
Proportion & $\phi$ &  $(1-\phi)$ \\\hline\hline
&&\\
Reject & $\alpha\phi$ & $(1-\beta)(1-\phi)$\\
&&\\\hline
&&\\
Not Reject & $(1-\alpha)\phi$ &  $\beta(1-\phi)$\\
&&\\\hline
\end{tabular}
\caption{Table showing the relative proportion of null hypotheses falling under each possible combination of true/false and reject/not reject for a family of statistical tests with Type-I error rate $\alpha$, Type-II error rate $\beta$, and prior odds $(1-\phi)/\phi$. }
\label{table:nohack}
\end{table}

The NHST protocol is prone to two types of error:
\begin{itemize}
	\item Type-I error: rejecting $H_0$ when $H_0$ is true; and
	\item Type-II error: failing to reject $H_0$ when $H_0$ is false.
\end{itemize} 
Since, when $H_0$ is true, the corresponding P-value is distributed uniformly on $(0,1)$, the probability of a Type-I error (i.e., obtaining `$P<\alpha$' when $H_0$ is true) is $\alpha$.  (In particular, when $\alpha=0.05$, the probability of Type-I error in any given test is 0.05.)
For a given Type-I error probability, the Type-II error rate, denoted as $\beta$, is the probability of failing to reject a false null hypothesis.  
The {\em power} is the probability $1-\beta$ of correctly rejecting a false null hypothesis.
Though norms vary among disciplines, it is conventional in many fields to tradeoff between Type-I and Type-II error by aiming for 80\% power at the 5\% significance level.  We use these as benchmarks in the empirical analysis below (see Section \ref{section:effects}).

\subsection{False positive rate}\label{section:FPR}

For a collection of statistical tests, the false positive rate (FPR) is the proportion of significant P-values (with $P<\alpha$) obtained under a true null hypothesis.  Since FPR reflects the proportion of significant P-values obtained in a collection of hypothesis tests, it depends on the proportion $\phi$ of all tests for which $H_0$ is true, in addition to the Type-I and Type-II error rates.  (The proportion $\phi$ is sometimes quoted in terms of the {\em prior odds $(1-\phi)/\phi$ in favor of $H_1$}, which is the relative proportion of false to true null hypotheses among all those tested.)
For tests conducted under the standard protocol with Type-I and Type-II error probabilities $\alpha$ and $\beta$ and prior odds $(1-\phi)/\phi$, the {\bf false positive rate} is given by
\begin{equation}\label{eq:FPR-nohack}\FPR(\alpha,\beta,\phi)=\frac{\alpha\phi}{\alpha\phi+(1-\beta)(1-\phi)}.\end{equation}
These quantities are summarized in Table \ref{table:nohack}.

\subsection{Replication Rate}

\begin{figure}[t]
\begin{center}
\scalebox{.5}{\includegraphics{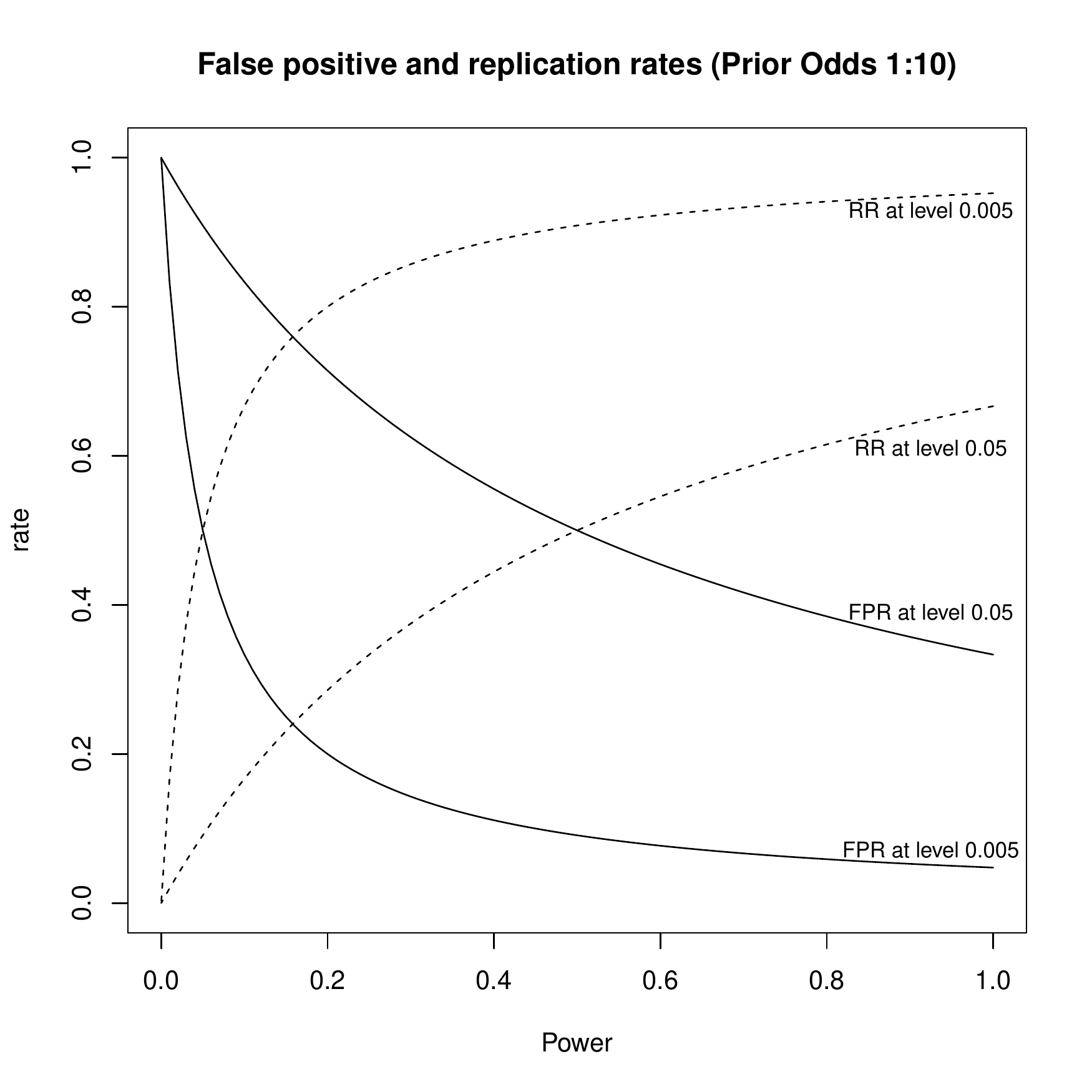}}
\end{center}
\caption{Plot of false positive rate (solid) and replication rate (dashed) at different levels of power for prior odds $1/10$ at significance levels 0.05 and 0.005.  The plot illustrates the complementary relationship between FPR and RR under the perfect reproducibility assumption, as shown in \eqref{eq:reciprocal}.}
\label{fig:FPR-RR-nohack}
\end{figure}

We consider a significant result ($P<\alpha$) to be {\em reproducible} (or {\em replicable}) if it is verifiable by subsequent testing.
Attempts to replicate often appeal to the same or similar statistical methods as the initial study, and are thus prone to the same sources of statistical error.
For a precise analysis of the replication rate, we rule out the possibility of Type-I and Type-II error in replication attempts and assume that 100\% of true positives and 0\% of false positives are replicable.\footnote{As a practical matter, this assumption is reasonable because replication studies typically seek to achieve high power ($>90\%$) at the 0.05 significance level, minimizing the occurrence of false positives and false negatives in replication attempts.}
This assumption, which we call {\bf perfect reproducibility}, is in the spirit of the reproducibility discussion in that it treats replication as a property of the finding itself: if the finding is true, then it is replicable; if it is false, then it is not.  
Under perfect reproducibility, the {\bf replication rate} for a family of tests with significance level $\alpha$, power $1-\beta$, and prior odds $(1-\phi)/\phi$ is the proportion of true positives,
\begin{equation}\label{eq:RR-nohack}
\RR(\alpha,\beta,\phi)=\frac{(1-\beta)(1-\phi)}{\alpha\phi+(1-\beta)(1-\phi)}.
\end{equation}
Comparing \eqref{eq:FPR-nohack} and \eqref{eq:RR-nohack}, we observe the complementary relationship between replication rate and false positive rate,
\begin{equation}\label{eq:reciprocal}\RR(\alpha,\beta,\phi)=1-\FPR(\alpha,\beta,\phi).\end{equation}
See Figure \ref{fig:FPR-RR-nohack} for a plot of FPR and RR against power for significance levels $0.05$ and $0.005$.

\section{NHST with P-hacking}\label{section:hack}

When analyzing the impact of P-hacking on replication, it is important to distinguish between the {\em protocol} of NHST and the {\em policy} for assigning the label of `statistical significance' to small P-values.  In a sound application of NHST, as assumed in Section \ref{section:nohack}, the protocol used to obtain the P-value is independent of the {policy} for conferring statistical significance.  
The calculations of false positive rate and replication rate in Section \ref{section:nohack} are thus valid in a world in which all P-values are obtained independently of the prevailing cutoff, as the NHST paradigm prescribes.  In such a world, all Type-I errors occur purely by chance, with probability $\alpha$, and both FPR and RR can be substantially improved by decreasing the significance level while holding power fixed, or alternatively by increasing power while holding significance level fixed, as Figure \ref{fig:FPR-RR-nohack} indicates.  
That world is a far cry from the world in which we live.

In the real world, many (and perhaps most) Type-I errors are the result of systematic departures from the NHST protocol which cause the observed false positive rate to be much larger than predicted by \eqref{eq:FPR-nohack}.  (These systematic departures are referred to generically as `P-hacking' here.)
 As a consequence, the standard theory of Section \ref{section:nohack} can no longer be applied to assess the false positive rate and replication rate in the presence of P-hacking. 
 
 To account for the effects of P-hacking, we distinguish between {\em sound} or {\em unsound} P-values:
\begin{itemize}
	\item A {\bf sound P-value} is one for which the standard interpretation is valid (i.e., the probability, under the null hypothesis, that the test statistic attains a value as extreme or more extreme than observed).
	\item An {\bf unsound} (or {\bf hacked}) {\bf P-value} is one which does not warrant the above interpretation because proper statistical protocol was not followed.  
\end{itemize}
For our purposes here, neither the specific type of P-hacking (e.g., multiple testing, selective reporting, etc.)~nor whether P-hacking was intentional is relevant.  What matters is (i) P-values obtained by P-hacking do not warrant the same interpretation as those obtained under the standard paradigm and (ii) since P-hacking protocol depends on the policy (i.e., cutoff value), the effect of a new policy on hacked P-values cannot be studied empirically based on data obtained under the present policy.
When considering the likely behavior of hacked P-values under the new cutoff, we cannot ignore that the scientific teams behind these P-values have been successful both in obtaining a significant P-value at the current significance level and in justifying why their hacked P-value gives a finding worthy of publication.  When the significance threshold changes, we can expect that these same scientists will still be successful in finding ways to obtain significant, publishable results. 
So, whereas the behavior of sound P-values can be treated as {\em absolute}---by virtue of being sound their behavior is unaffected by policy changes---the behavior of unsound P-values must be interpreted {\em relative} to the prevailing policy.\footnote{In making this observation, I do not mean to represent P-hacking as a universally malicious, subversive activity by which ne'er-do-well scientists try to make a P-value as small as possible at all costs.  In many cases, scientists have been trained that what we are here calling P-hacking is a legitimate research technique.  I discuss practical matters of this sort in Section \ref{section:concluding}.}
For example, a hacked P-value of 0.045 under the current 0.05 significance regime should not be expected to remain at 0.045 after the cutoff changes to 0.005.  By virtue of being hacked, this P-value is already smaller than it ought to be, as a consequence of so-called `researcher degrees of freedom'.  If the significance policy were changed to 0.005, this same P-value should be expected to decrease, and would likely end up below 0.005 upon further application of said degrees of freedom.

\subsection{False positive rate and replication rate under P-hacking}

The classification of P-values as sound and unsound identifies two different regimes to consider when computing FPR.  Of all computed P-values, we write $0<h<1$ to denote the proportion obtained by unsound protocol (i.e., not according to the theory of Section \ref{section:nohack}).  With $0<\alpha<1$ fixed as a baseline significance level, we assume that this entire proportion $h$ is significant at level $\alpha$ {\em when the prevailing significance cutoff is $\alpha$}.\footnote{In making this assumption, we ignore attempts at P-hacking which fail to obtain a P-value less than $\alpha$.  Since these P-values are never labeled `significant', they are assumed not to be observed in the scientific literature and therefore do not contribute to the reproducibility crisis. } 
The remaining proportion $1-h$ of P-values are sound and behave according to the analysis in Section \ref{section:nohack}.

\begin{table}[t]
\begin{tabular}{|c||c|c|c|}
\hline
&\multicolumn{2}{|c|}{ }&\\
& \multicolumn{2}{|c|}{$H_0$ true} & $H_0$ false\\
&\multicolumn{2}{|c|}{ }&\\\hline
& Sound & Unsound & Sound\\
Proportion & $\phi(1-h)$ & $h$ & $(1-\phi)(1-h)$ \\\hline\hline
&&&\\
Reject & $\alpha(1-h)\phi$ & $h$ & $(1-\beta)(1-h)(1-\phi)$\\
&&&\\\hline
&&&\\
Not Reject & $(1-\alpha)(1-h)\phi$ & $0$ & $\beta(1-h)(1-\phi)$\\
&&&\\\hline
\end{tabular}
\caption{Table showing the relative proportion of null hypotheses corresponding to true/false positives/negatives in the presence of P-hacking.  The proportion $1-h$ of sound tests has Type-I error rate $\alpha$, Type-II error rate $\beta$, and prior odds $(1-\phi)/\phi$. Table assumes $\alpha$ as the baseline significance level, so that 100\% of the proportion $h$ of hacked P-values are significant at level $\alpha$.  By setting $h=0$, these proportions coincide with those in Table \ref{table:nohack}.
}
\label{table:hack}
\end{table}

The proportions of significant P-values of each type (i.e., sound true positive, sound false positive, and unsound false positive) for a given choice of $\beta$ and $\phi$ are now
\begin{align*}
(1-\beta)(1-h)(1-\phi)&: \text{ sound true positive}\\
\alpha (1-h)\phi&: \text{ sound false positive}\\
h&: \text{ unsound false positive}.
\end{align*}
See Table \ref{table:hack} for a breakdown of these proportions.\footnote{Since the primary contribution to the reproducibility crisis is the disproportionate number of false positives obtained by P-hacking, we also rule out the possibility of obtaining a `true positive' via P-hacking.  Under perfect reproducibility, such findings would be reproducible, and thus would not contribute to reproducibility issues.   }  
With the addition of $h$, the false positive rate is now computed as
\begin{equation}\label{eq:FPR}
\FPR(\alpha,\beta,\phi;h)=\frac{\alpha(1-h)\phi+h}{\alpha(1-h)\phi+(1-\beta)(1-h)(1-\phi)+h}.
\end{equation}
Under the assumption of perfect reproducibility, we once again observe the complementary relationship between false positive rate and replication rate
\begin{equation}\label{eq:RR}\RR(\alpha,\beta,\phi;h)=1-\FPR(\alpha,\beta,\phi;h).\end{equation}
By setting $h=0$ we immediately recover \eqref{eq:FPR-nohack}, thus observing the precise sense in which the calculations in Section \ref{section:nohack} ignore P-hacking.

\subsection{P-value distributions}

To analyze how each of the three classes of significant P-value at level $\alpha$ will behave upon lowering the cutoff to $\alpha/c$, for $c\geq1$, we treat the behavior of sound P-values as {\em absolute}---sound P-values do not depend on the policy used in determining significance---and unsound P-values as {\em relative}---in the absence of other information, an unsound P-value satisfying $P<\alpha$ when the cutoff is  $\alpha$ should also be expected to satisfy $P<\alpha/c$ when the cutoff is changed to $\alpha/c$. 
  The key point is that hacked P-values are not hacked specifically to be below {\em $\alpha$}.  They are hacked to be below the {\em significance threshold}, which just so happens to be $\alpha$.  When the threshold changes, so does the target, and likely so will the P-value.

A key consideration when assessing the impact of a regime change is the extent to which hacked P-values under one regime will persist (i.e., continue to be significant) under the new regime.  
We write $\psi_{\alpha/c}(x)$, $0<x<1$, to denote the proportion of hacked P-values in the range $(0,x)$ when the significance cutoff is $\alpha/c$, for $c\geq1$.  Since a hacked P-value is not expected to increase in value upon lowering the cutoff, we assume that these distributions are monotone with respect to significance level, in the sense that a hacked P-value which is less than $x$ at level $\alpha$ will also be less than $x$ if the cutoff is decreased to $\alpha'<\alpha$.  Thus, we assume $\psi_{\alpha/c}(x)\geq\psi_{\alpha}(x)$ for all $c\geq1$ and $0<x<1$.  In particular, by setting the baseline significance level at $\alpha$, we assume $\psi_{\alpha/c}(\alpha)=1$ for all $c\geq1$, and $h\psi_{\alpha}(\alpha)=h$ is the proportion of all P-values that are unsound and significant at level $\alpha$.

For $c\geq1$, $\psi_{\alpha/c}(\alpha/c)$ is the proportion of all hacked P-values which are significant when the significance threshold is decreased to $\alpha/c$.  The proportion $\psi_{\alpha/c}(\alpha/c)$ is central to our assessment of the RSS proposal, as it quantifies the extent to which hacked P-values that are significant at level $\alpha$ will persist (and remain significant) under the lower $\alpha/c$ cutoff.
The analysis in \cite{RSS} compares the empirical replication rate for P-values with $P<0.005$ and $0.005<P<0.05$ observed in \cite{Camerer2016,PsychRep} to arrive at their suggested 2-to-1 increase in replication rate under the lower cutoff. 
But since the behavior of hacked P-values depends on the cutoff, we cannot naively estimate $\psi_{\alpha/c}(\alpha/c)$ 
by $\psi_{\alpha}(\alpha/c)$ (i.e., the proportion of hacked P-values that lie below $\alpha/c$ when the cutoff is $\alpha$).  
Instead, we should expect $\psi_{\alpha/c}(\alpha/c)$ to lie somewhere between $\psi_{\alpha}(\alpha/c)$ and $\psi_{\alpha}(\alpha)$, reflecting the inevitability that P-hackers will adapt to the new significance regime.

Since $\psi_{\alpha/c}(\alpha/c)$ depends on the cutoff $\alpha/c$, we cannot estimate it from data observed under the current $\alpha$ cutoff.  When analyzing the potential impact of the lower cutoff, we consider all possible values in the range $\psi_{\alpha}(\alpha/c)\leq\psi_{\alpha/c}(\alpha/c)\leq\psi_{\alpha}(\alpha)$ by interpolation,
\begin{equation}\label{eq:pi}\psi_{\alpha/c}(\alpha/c)=\pi\psi_{\alpha}(\alpha)+(1-\pi)\psi_{\alpha}(\alpha/c),\end{equation}
for $0<\pi<1$.
The parameter $\pi$ quantifies the rate at which hacked P-values `persist' after a change in cutoff.  We therefore call $\pi$ the {\em persistence parameter} and sometimes refer to $\psi_{\alpha/c}(\alpha/c)$ simply as the {\em persistence} at level $\alpha/c$.
Note that $\pi=1$ corresponds to maximal persistence (i.e., $\psi_{\alpha/c}(\alpha/c)=\psi_{\alpha}(\alpha)=1$), so that 100\% of hacked significant P-values under the current cutoff are significant at the new cutoff; and $\pi=0$ corresponds to minimal persistence (i.e., $\psi_{\alpha/c}(\alpha/c)=\psi_{\alpha}(\alpha/c)$) so that only those hacked P-values that currently lie below $\alpha/c$ are significant at the new threshold.

\subsection{False positive and replication rates under regime change}

\begin{table}[t]
\begin{tabular}{|c||c|c|c|}
\multicolumn{4}{c}{Under $\alpha/c$ cutoff}\\
\multicolumn{4}{c}{ }\\
\hline
&\multicolumn{2}{|c|}{ }&\\
& \multicolumn{2}{|c|}{$H_0$ true} & $H_0$ false\\
&\multicolumn{2}{|c|}{ }&\\\hline
& Sound & Unsound & Sound\\
Proportion & $\phi(1-h)$ & $h$ & $(1-\phi)(1-h)$ \\\hline\hline
&&&\\
Reject & $\alpha(1-h)\phi$ & $h\psi_{\alpha/c}(\alpha/c)$ & $(1-\beta)(1-h)(1-\phi)$\\
&&&\\\hline
&&&\\
Not Reject & $(1-\alpha)(1-h)\phi$ & $h(1-\psi_{\alpha/c}(\alpha/c))$ & $\beta(1-h)(1-\phi)$\\
&&&\\\hline
\end{tabular}
\caption{Table showing the relative proportion of null hypotheses corresponding to true/false positives/negatives in the presence of P-hacking with persistence $\psi_{\alpha/c}(\alpha/c)$ at level $\alpha/c$.  The proportion $1-h$ of sound tests has Type-I error rate $\alpha$, Type-II error rate $\beta$, and prior odds $(1-\phi)/\phi$. }
\label{table:hack-persistent}
\end{table}

In taking $\alpha$ as the baseline, we set $\psi_{\alpha}(\alpha)=1$ so that the proportion of unsound significant P-values at level $\alpha$ equals $h\psi_{\alpha}(\alpha)=h$ as in Table \ref{table:hack}.
Assuming $h$ is independent of the cutoff, the proportion of all hacked P-values that are significant at the new $\alpha/c$ cutoff is given by $\psi_{\alpha/c}(\alpha/c)$, as reflected in the updated proportions of Table \ref{table:hack-persistent}.
The cutoff-dependent distribution $\psi$ of unsound P-values now features in the calculation of false positive rate and replication rate from \eqref{eq:FPR} and \eqref{eq:RR} by
\begin{align}
\label{eq:FPR-hack-persistent}
\FPR(\alpha/c,\beta,\phi;h,\psi)&=\frac{(\alpha/c)\phi(1-h)+h\psi_{\alpha/c}(\alpha/c)}{(\alpha/c)\phi(1-h)+h\psi_{\alpha/c}(\alpha/c)+(1-\beta)(1-\phi)(1-h)}\quad\text{and}\\
\label{eq:RR-hack-persistent}
\RR(\alpha/c,\beta,\phi;h,\psi)&=\frac{(1-\beta)(1-\phi)(1-h)}{(\alpha/c)\phi(1-h)+h\psi_{\alpha/c}(\alpha/c)+(1-\beta)(1-\phi)(1-h)}.
\end{align}
The assumed form of $\psi_{\alpha/c}$ in \eqref{eq:pi} and the convention $\psi_{\alpha}(\alpha)=1$ implies $\psi_{\alpha/c}(\alpha/c)\geq\pi$, giving the bounds
\begin{align}
\FPR(\alpha/c,\beta,\phi;h,\psi)&\geq\frac{(\alpha/c)\phi(1-h)+h\pi}{(\alpha/c)\phi(1-h)+h\pi+(1-\beta)(1-\phi)(1-h)}\label{eq:FPR-bound}\\
\RR(\alpha/c,\beta,\phi;h,\psi)&\leq\frac{(1-\beta)(1-\phi)(1-h)}{(\alpha/c)\phi(1-h)+h\pi+(1-\beta)(1-\phi)(1-h)},\notag
\end{align}
which we use to obtain conservative estimates in the forthcoming empirical analysis.

\section{Analysis of ``Redefining Statistical Significance''}\label{section:effects}

For the sake of this analysis, we consider the potential effects of changing the baseline significance level from $\alpha=0.05$ to $\alpha/c=0.005$.  
We take 0.80 as the conventional level of power (i.e., $\beta=0.20$) at significance level $0.05$.
We assume prior odds of $1/10$ for results in psychology, as suggested by empirical evidence in that field \cite{Johnson2016}.
Since the analysis of FPR (Equation \eqref{eq:FPR-hack-persistent}) under the change in cutoff relies on the persistence parameter $\pi$  which is not estimable from data, we present all possible values of FPR over the entire range $0<\pi<1$ whenever applicable. 
  Although it is possible that the prior odds could change in response to redefining statistical significance, 
  we assume for simplicity that $\phi$ will remain unchanged upon lowering the cutoff to 0.005.  

\subsection{Estimating the hacking rate}

In showing the sensitivity of FPR (and therefore RR) to the rate of P-hacking ($h$), Figure \ref{fig:P-hacking}  
immediately raises doubts about the analysis in \cite{RSS}.  The expressions of FPR and RR in \eqref{eq:FPR-hack-persistent} and \eqref{eq:RR-hack-persistent}, which account for P-hacking, allow us to examine these doubts and arrive at the conclusions highlighted in the abstract and introductory section.
In particular, for $0.05<h<0.15$, we see that the FPR is much higher than suggested by the analysis in \cite{RSS}, and any improvement due to lowering the cutoff will be minimal.  To validate this observation, we estimate the hacking rate based on data from 
 a recent replication project in psychology \cite{PsychRep}.

We obtain a range of estimates for $h$ by comparing the empirical observations about replication obtained in \cite{PsychRep} with the predicted FPR and RR under P-hacking in \eqref{eq:FPR} and \eqref{eq:RR}.  The replication study in \cite{PsychRep} shows an empirical replication rate of $37\%$ (36 out of 97) in the field of psychology, far below the rate of 62\% predicted by \eqref{eq:RR-nohack} with $\alpha=0.05$, $\beta=0.20$, and prior odds $10/11$.  Fitting this observed value to \eqref{eq:RR} gives a point estimate of $h=0.075$.  By further comparing the observed replication rate among P-values which were originally in the ranges $P<0.005$ (24 out of 47) and $0.005<P<0.05$ (12 out of 50), we obtain estimates for $h$ ranging between $0.05$ and $0.15$.
  These estimates are consistent with the widespread belief that P-hacking is prevalent in the psychology literature.\footnote{These estimates, which were obtained merely by matching theoretical and empirical values of replication rate based on a single study, are not put forward here as estimates of the `real' rate of P-hacking in psychology or any other discipline.   They are intended merely to provide a ballpark figure for the sake of analyzing the RSS proposal in light of P-hacking. }

\subsection{The impact of redefining statistical significance}\label{section:Benjamin}

Benjamin, et al cite two main pieces of evidence in support of their proposal to change the significance cutoff from 0.05 to 0.005:
\begin{itemize}
	\item[(i)] The observation (partially reconstructed in Figure \ref{fig:P-hacking}) that the false positive rate is unacceptably high (greater than 0.33 for all levels of power) under the current 0.05 cutoff, and that these rates will fall to more acceptable levels (below 0.10 for many levels of power) under the new 0.005 cutoff.  Compare the solid lines in Figure \ref{fig:P-hacking}, also \cite[Figure 2]{RSS}.
	\item[(ii)] The empirical studies of replication rates in psychology \cite{PsychRep} and experimental economics \cite{Camerer2016} suggest that the replication rate for findings with P-value in the range $P<0.005$ is twice that of P-values in the complementary range $0.005<P<0.05$.
\end{itemize}
Neither claim is valid.
 Claim (i) relies on theory (Section \ref{section:nohack}) which tacitly ignores P-hacking, and thus does not apply in the context for which the RSS proposal was designed.
 Claim (ii) ignores the dependence of $\psi_{\alpha}$ on the significance cutoff $\alpha$, as supported by empirical studies \cite{Head2015,Mariscampo2012,Simonsohn2014} and common sense (i.e., as P-hackers currently flout the prescribed protocol, knowingly or unknowingly, they should be expected to continue doing so when the prescription changes).
 With the dependence of $\psi_{\alpha}$ on the significance cutoff accounted for, we see that the suggested benefits of lowering the significance threshold (i.e., FPR decreases by a factor greater than two and replication rate doubles) are far less certain than presented in \cite{RSS}.  Quite simply, such projections are incongruous with reality. 

\begin{figure}[t]
\begin{center}
\scalebox{.4}{\includegraphics{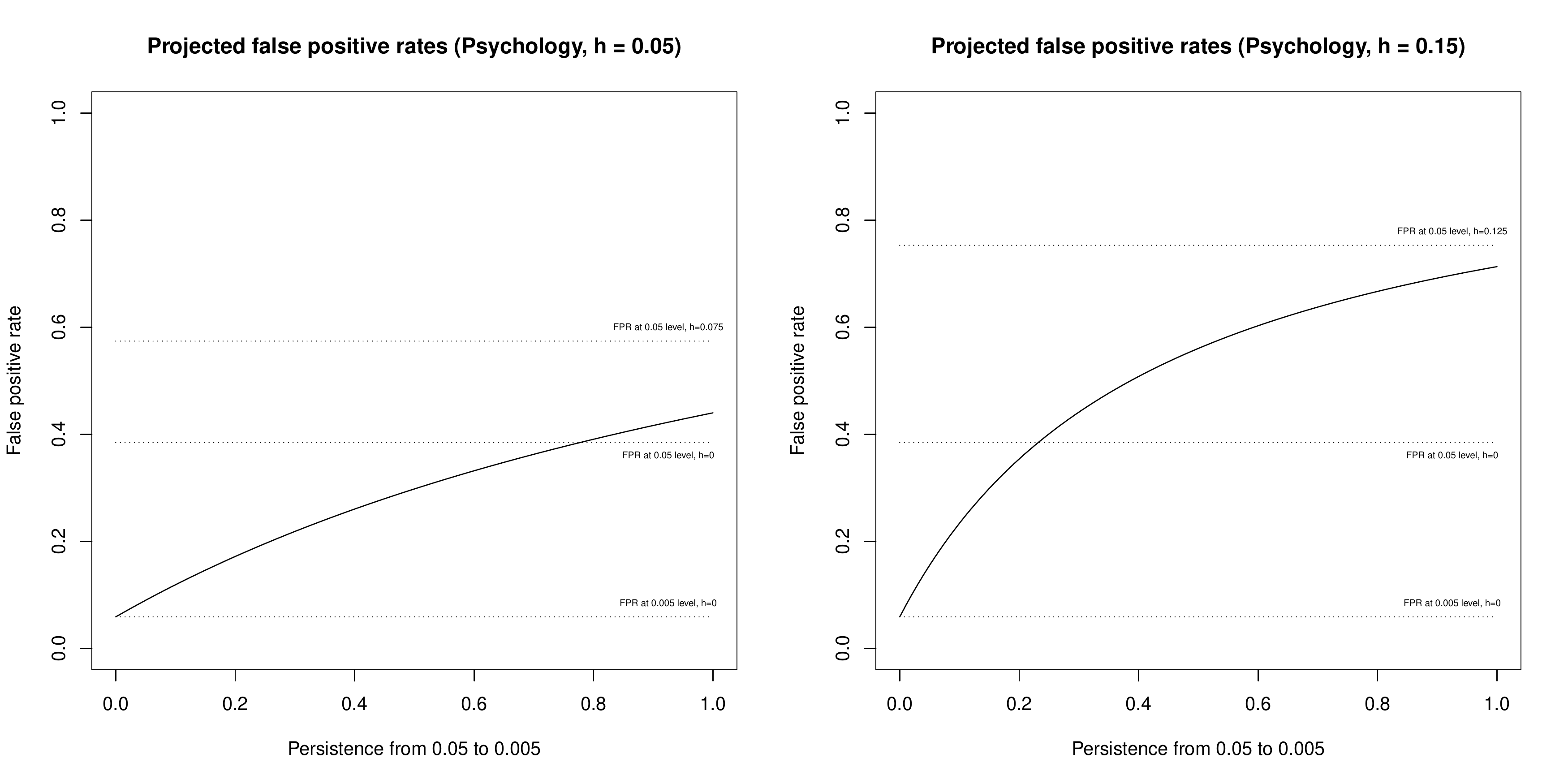}}
\end{center}
\caption{Solid lines show the projected false positive rates for $h=0.05$ (left) and $h=0.15$ (right) over the full range $0<\pi<1$ of persistence parameters.  Both plots assume $\alpha=0.005$, $\beta=0.20$, and prior odds $1/10$.  The plot on the left assumes $h=0.05$ and the plot on the right assumes $h=0.15$.  The dotted lines are the false positive rates computed under the estimated value of $h$ (top line) and by setting $h=0$ for significance levels $0.05$ (middle) and $0.005$ (bottom). }
\label{fig:FPR-psi}
\end{figure}

\subsubsection*{Claimed benefits of the RSS proposal are exaggerated}

Figure \ref{fig:FPR-psi} shows how the false positive rate will change (as a function of persistence $\pi$) compared to several reference points.  
The top dotted line in both figures is the FPR for $h=0.05$ and $h=0.15$ at the 0.05 significance level.  In both cases, the FPR is much higher than the standard theory predicts in the absence of P-hacking (as given by the middle dotted line in both plots).  The bottom dotted line is the FPR predicted at significance level 0.005 in the absence of P-hacking.  The solid curve shows how FPR varies (at the 0.005 cutoff) with the persistence $\pi$. 
To appreciate the discrepancy between what can be expected in the presence of P-hacking and what is claimed in \cite{RSS}, note the difference between the solid lines in Figure \ref{fig:FPR-psi} and the bottom dotted line in each plot, which represents the predicted FPR purported by the RSS analysis. 
Notice that FPR under the new 0.005 cutoff will remain above 20\% even if persistence is relatively low ($\pi\approx0.25$) and power is assumed to remain at 0.80 under the new cutoff.  For high levels of persistence (e.g., $\pi\geq0.50$), the FPR is near or far above what would be expected under the 0.05 level in the absence of P-hacking.  

\begin{figure}[t]
\begin{center}
\scalebox{.4}{\includegraphics{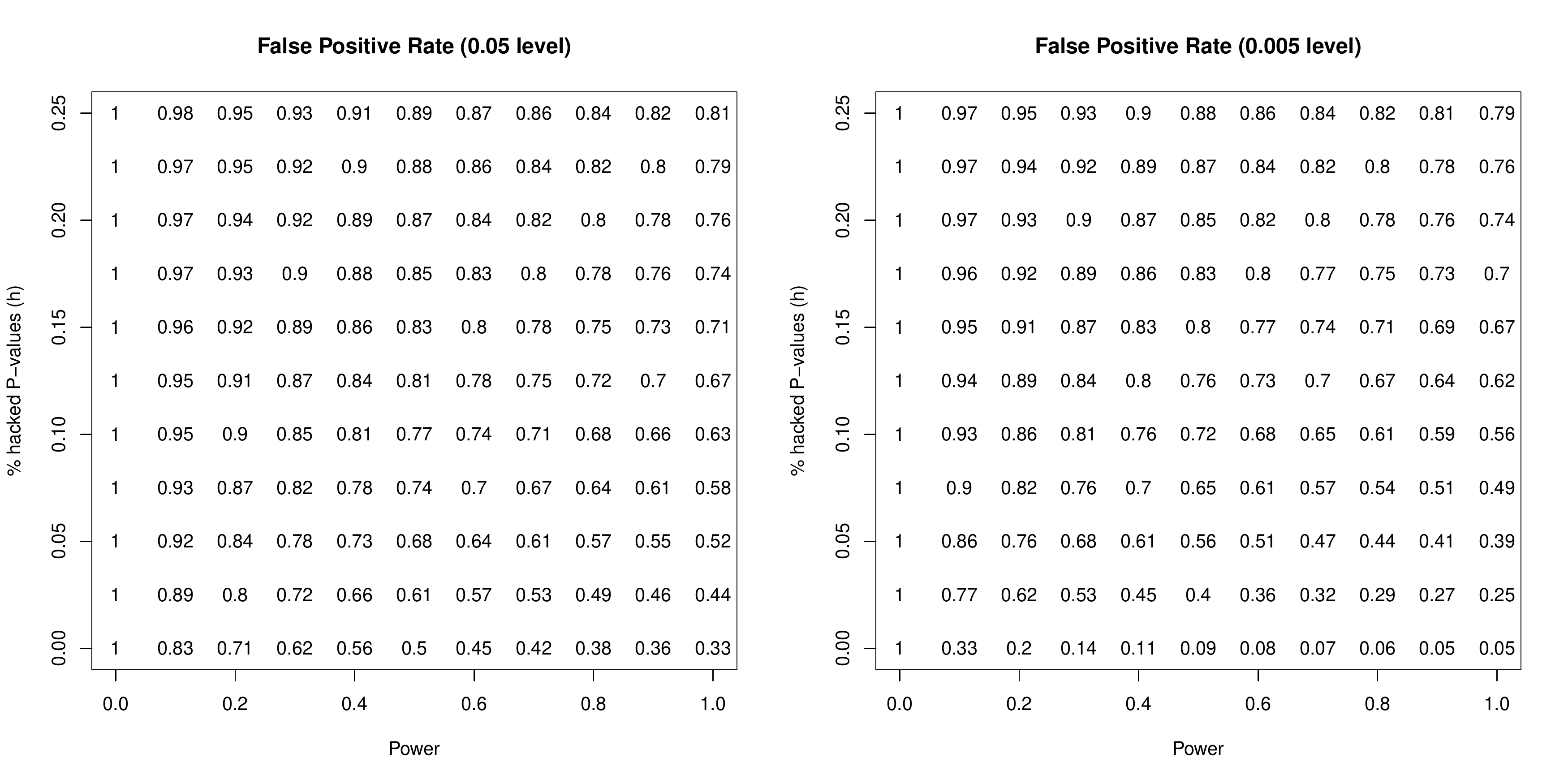}}
\end{center}
\caption{False positive rates under different combinations of power ($1-\beta$) and P-hacking rate ($h$) for significance level 0.05 (left) and 0.005 (right).  The prior odds is taken to be 1:10 (i.e., $\phi=10/11$) in both cases and power at the 0.05 level fixed at 0.80.}
\label{fig:power-h}
\end{figure}

\subsubsection*{False positive rate will remain high under RSS}

A finer grained analysis is given in Figure \ref{fig:power-h}, which plots FPR under different combinations of power and persistence $\psi_{0.005}(0.005)$.
We note that the analysis put forward by RSS in support of redefining the cutoff from $0.05$ to $0.005$ assumes $h=0$, which for a power of 0.80 leads to a decrease in FPR from 0.38 (under the 0.05 cutoff) to 0.06 (under the 0.005 cutoff).  
In the absence of P-hacking, the perceived improvement is appealing in both relative (decrease by factor greater than 6) and absolute (decrease false positive rate to 6\%) terms.  However, the improvements are substantially mitigated in the presence of even modest P-hacking: for $h=0.05$ and power of 0.80, FPR decreases from 0.57 to 0.44; and for $h=0.15$ and power 0.80, FPR decreases from 0.75 to 0.71.  So if power can be maintained after the significance cutoff is decreased, then the false positive rate will go down, but by much more modest factors (between $5\%$ and $22\%$) which are unlikely to be noticeable in practice and would still leave FPR at much higher levels than desired.

\subsubsection*{The reproducibility crisis could get worse}

So far we have granted the RSS proposal the benefit of the doubt in assuming that power of 0.80 can be maintained after the change in cutoff.  This is possible, but not without additional costs (i.e., time and money) due to the need of larger sample sizes.
Even granting this benefit, the unimpressive improvements to FPR shown in Figures \ref{fig:FPR-psi} and \ref{fig:power-h} call into question the RSS assertion that the benefits of the lower cutoff would outweigh the costs of achieving the same level of power.  Furthermore, the practical difficulty of achieving such high power without compromising other parts of the study cannot be ignored in fields such as psychology, where the reproducibility crisis seems most pronounced.
It seems inevitable that the power of many studies would decrease under the new cutoff.

\begin{figure}[t]
\begin{center}
\scalebox{.4}{\includegraphics{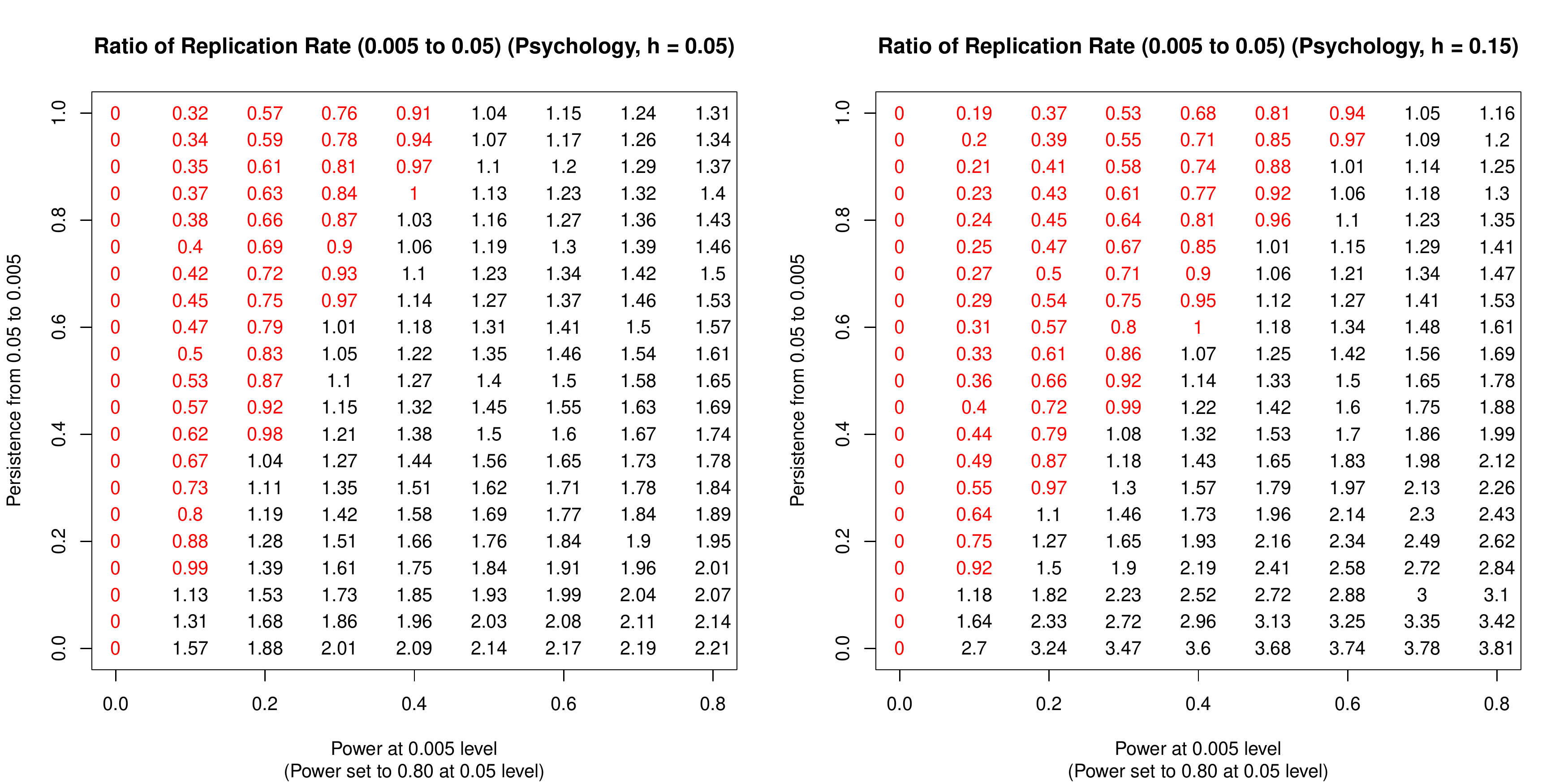}}
\end{center}
\caption{Ratio of RR under 0.005 and 0.05 cutoffs based on the psychology replication data.  (Left) Shows ratio 
\[\frac{\RR(0.005,\beta,10/11;0.05,\psi_{0.005}(0.005))}{\RR(0.05,0.2,10/11;0.05,1)}\]
for different combinations of power $1-\beta$ and persistence $\psi_{0.005}(0.005)$. (Right) Shows ratio \[\frac{\RR(0.005,\beta,10/11;0.15,\psi_{0.005}(0.005))}{\RR(0.05,0.2,10/11;0.15,1)}\]
 for different combinations of power $1-\beta$ and persistence $\psi_{0.005}(0.005)$.  Ratios less than 1 are colored in red, indicating that replication rate is worse under the lower 0.005 cutoff. }
\label{fig:FPR-ratio}
\end{figure}

To examine the impact of a potential loss in power, 
Figure \ref{fig:FPR-ratio} plots the ratios of replication rate for different choices of power ($1-\beta$) and persistence ($\psi_{0.005}(0.005)$) to the replication rate under the 0.05 level with power 0.80.  Ratios smaller than 1 are colored red to indicate that the replication rate gets worse under the lower cutoff of 0.005.
As expected, the replication rate improves as long as the same level of 0.80 power can be maintained at the lower cutoff.  But only under special circumstances does the ratio exceed 2, as the authors suggest based on the empirical evidence in \cite{Camerer2016} and \cite{PsychRep}.  For the replication rate to double, the persistence can be no greater than 0.35 (at power 0.80 and $h=0.15$) and no greater than 0.15 (at power 0.80 and $h=0.05$).   Based on these observations, it is hard to imagine a circumstance under which the replication rate would improve by a factor greater than $1.5$.  

Since the RSS proposal takes no active step to combat the root cause of the reproducibility crisis (i.e., P-hacking), it is unrealistic to expect P-hacking to improve in any noticeable way under the new proposal.
And if the same level of power cannot be maintained, then the situation is even more uncertain, with most scenarios suggesting modest gains or even losses in replication rate.
For example, if power falls to 0.50 under the new 0.005 cutoff and P-hacking persists at rate greater than 0.75, then replication rate could increase by as much as 19\% (at persistence $0.75$ and $h=0.05$) or decrease by 19\% (at persistence 1 and $h=0.15$).  The replication rate under the new cutoff would vary between 20\% and 51\% in such a case, as can be compared to the 37\% replication rate observed in \cite{PsychRep}.

\subsection{How persistent are P-hackers?}
A determining factor in evaluating the impact of the RSS proposal is the extent to which the lower cutoff will lead to a reduction in the proportion of hacked P-values that attain statistical significance.   
Since RSS takes no active steps to deter P-hacking, it is unlikely to have any positive impact on the prevalence of P-hacking.  In particular, we cannot rule out the possibility that $\psi_{0.005}(0.005)\approx1$, in which case the benefit to reproducibility would be undetectable or even negative if the same level of power cannot be achieved: see the top line in Figure \ref{fig:FPR-ratio}.
Without compelling evidence to the contrary, we should expect P-hacking to continue just as it is currently.
The authors of RSS have tried to preempt this criticism, arguing that ``[r]educing the P-value
threshold complements---but does not substitute for---solutions to these other problems'' \cite{RSS}.  Based on our above analysis, however, we find little `complementary' about the RSS proposal.  All available data points to a marginal improvement in FPR as long as the lower significance level has no adverse effect on the level of statistical power ($1-\beta$), hacking rate ($h$), or prior odds ($(1-\phi)/\phi$).  This alone is a strong assumption.  Whatever improvements to reproducibility might result from the other efforts to thwart P-hacking will have been achieved in spite of the RSS proposal.

\section{Concluding remarks}\label{section:concluding}

Using the same theoretical device (i.e., false positive rate under NHST) and empirical evidence (i.e., the psychology replication study in \cite{PsychRep}), we have analyzed the RSS proposal in light of claims that it will improve reproducibility.  By accounting for the effects of P-hacking, we see that the claimed benefits to false positive rate and replication rate are much less certain than suggested in \cite{RSS}.
In fact, if false positive rate were to decrease at all, it will be virtually unnoticeable, and will remain much higher than claimed in \cite{RSS}.  Replication rates will not come close to doubling unless the lower cutoff successfully eliminates all but a small fraction of P-hacking.  All available evidence suggests that P-hacking would continue much as it is now, and replication rates would either improve marginally or even decrease and would remain far below what is necessary to improve reproducibility in any substantive way.
Altogether, these observations point to one conclusion:
{ The proposal to redefine statistical significance is severely flawed, presented under false pretenses, supported by a misleading analysis, and should not be adopted.}

Defenders of the proposal will inevitably criticize this conclusion as ``perpetuating the {\em status quo},'' as one of them already has \cite{Machery2017}.  Such a rebuttal is in keeping with the spirit of the original RSS proposal, which has attained legitimacy not by coherent reasoning or compelling evidence but rather by appealing to the authority and number of its 72 authors.  The RSS proposal is just the latest in a long line of recommendations aimed at resolving the crisis while perpetuating the cult of statistical significance \cite{Cult} and propping up the flailing and failing scientific establishment under which the crisis has thrived.
The proposal to redefine statistical significance {\em is} the {\em status quo} made by a group of authors who {\em represent} the {\em status quo}:  advocates of the proposal not only embrace the bureaucracy of `statistical significance' but also contribute to the {\em status quo} replication crisis by peddling an argument grounded in unstated assumptions and incomplete analysis.
This latter point cannot be overstated: if the authors of this proposal, many of whom are prominent in their respective fields and have been active in combatting the reproducibility crisis for years, are prone to the same fundamental errors that are largely responsible for the replication crisis, then what hope is there to stem the crisis in the scientific communities they lead?

  While I am sympathetic to the sentiment prompting the various responses to RSS \cite{Amrhein2017, Lakens2017,McShane2017,Trafimow2017}, I am not optimistic that the problem can be addressed by ever expanding scientific regulation in the form of proposals and counterproposals advocating for pre-registered studies, banned methods, better study design, or generic `calls to action'. 
Those calling for bigger and better scientific regulations ought not forget that another regulation---the 5\% significance level---lies at the heart of the crisis.  

Beyond their implications for the specific recommendation to redefine statistical significance, the subtle but severe issues with the RSS proposal 
call into question the role of statistics in scientific research \cite{CraneMartin2017}.
The past century has seen the steady rise of statistical thinking as a method for validating scientific findings \cite{Porter198}.  Statistics is now the sole adjudicator of what is `significant' and what is not, what is scientific and what is not, what is important and what is not.
Other techniques have largely been phased out, slowly but steadily \cite{Gigerenzer1989,Porter200}, to such an extent that  it is now difficult, or impossible, in some fields to describe what constitutes `knowledge' without referring to the concept of statistical significance, or some similar statistical measure.   
Despite this, or perhaps because of it, science finds itself in the midst of crisis.

Viewed in this light, P-hacking and its consequences are seen for what they are: not as a defective or fraudulent behavior but rather as a natural response to overbearing and repressive regulation.
For the practicing scientist, statistics is just one of many potential ways to glean insights and validate findings.  While some hypotheses may be properly tested and understood by statistical reasoning, others might not be amenable to any formal quantitative or statistical method.  The rise of statistical thinking over the past century \cite{Gigerenzer1989,Porter198} has crowded out these other ways of arriving at truth, which despite being more qualitative, less precise, and more ambiguous are no less valid.  
In such a parochial system, what option other than P-hacking is available for the scientist who in earnest believes to have come upon a veritable finding, which despite having a sound basis in logic and evidence cannot be tested statistically?

I have commented before \cite{Crane2017Gelman} about the need to break free from the modern data-crazed mindset,  which fetishizes the `statistical' and brushes aside everything else.
 To borrow from Feyerabend \cite[p.\ 7]{Feyerabend}, ``The only principle that does not inhibit progress is: anything goes''. Scientists ought to be encouraged to make convincing, cogent arguments for their hypotheses however they see fit, without the decree of mandated protocol or embargoed methods.  
 
  To be clear, I am not calling for a `ban' of statistics---I am not calling for a ban of {\em anything}. 
The reproducibility crisis exposes the vulnerability of relying too heavily on any one paradigm, statistical or otherwise, and the folly of entrusting the arbitration of `knowledge' to an enlightened intellectual bureaucracy which keeps the proletariat in check by `raising awareness', banning methods, and redefining what is significant.
  These vulnerabilities are only magnified by the allure of publication, prestige, promotion, and the many other human (all too human) factors whose influence is undeniable and inevitable.   Against conventional wisdom, I contend that the reproducibility crisis cannot be fixed by better statistical education, increased awareness, patronizing `how to' guides (e.g., \cite{Kass}), or enhanced oversight by journals and intellectual bureaucrats.  
  These measures have been tried, and they have failed.
  Such steps only accelerate bureaucracy, which in turn calcifies the {\em status quo} and further promotes a collective inability to conceive of what constitutes `knowledge' independently of the bureaucratic policies used to evaluate `findings'.
The only way to reverse course is to loosen---not tighten---the restrictions on what makes an analysis {\em scientific} and a finding {\em significant}.

\end{document}